\documentclass[]{aa} 
\usepackage{natbib}
\usepackage[dvips, pdftex]{graphicx}
\usepackage{epstopdf}
\usepackage{txfonts}
\bibpunct{(}{)}{;}{a}{}{,}

\begin{document}
\authorrunning{Hu et al.}
\titlerunning{Impact of helium diffusion and helium-flash-induced carbon production on gravity-mode pulsations in sdB stars}

   \title {Impact of helium diffusion and helium-flash-induced carbon production on gravity-mode pulsations in subdwarf B stars}

   \author{Haili Hu
          \inst{1,2}        
          \and                                        
          G.~Nelemans\inst{1}
          \and                            
          C.~Aerts\inst{1,2}   
          \and
          M.-A.~Dupret\inst{3}                                                                                 
         }

   \offprints{hailihu@astro.ru.nl}

   \institute{Department of Astrophysics, IMAPP, Radboud University Nijmegen, P.O.~Box 9010, 6500 GL Nijmegen, The Netherlands
         \and
             Instituut voor Sterrenkunde, K.U.Leuven, Celestijnenlaan 200D, B-3001 Leuven, Belgium
         \and                
         Institute d'Astrophysique et G\'eophysique, universit\'e de Li\`ege, Belgium            
             }

   \date{Received 16 June 2009 / Accepted 1 October 2009 }

 \abstract
   {Realistic stellar models are essential to the forward modelling approach in asteroseismology. For practicality however, certain model assumptions are also required. For example, in the case of subdwarf B stars, one usually starts with zero-age horizontal branch structures without following the progenitor evolution. 
   }
   {We analyse the effects of common assumptions in subdwarf B models on the $g$-mode pulsational properties. We investigate if and how the pulsation periods are affected by the H-profile in the core-envelope transition zone. Furthermore, the effects of C-production and convective mixing during the core helium flash are evaluated. Finally, we reanalyse the effects of stellar opacities on the mode excitation in subdwarf B stars.}
   {We computed detailed stellar evolutionary models of subdwarf B stars, and their non-adiabatic pulsational properties. Atomic diffusion of H and He is included consistently during the evolution calculations. The number fractions of Fe and Ni are gradually increased by up to a factor of 10 around $\log T=5.3$. This is necessary for mode excitation and to approximate the resulting effects of radiative levitation. We performed a pulsational stability analysis on a grid of subdwarf B models constructed with OPAL and OP opacities.}
  {We find that helium settling causes a shift in the theoretical blue edge of the $g$-mode instability domain to higher effective temperatures. This results in a closer match to the observed instability strip of long-period sdB pulsators, particularly for $ l\leq3$ modes.  
   We show further that the $g$-mode spectrum is extremely sensitive to the H-profile in the core-envelope transition zone. If atomic diffusion is efficient, details of the initial shape of the profile become less important in the course of evolution. Diffusion broadens the chemical gradients, and results in less effective mode trapping and different pulsation periods. Furthermore, we report on the possible consequences of the He-flash for the $g$-modes. The outer edge of a flash-induced convective region introduces an additional chemical transition in the stellar models, and the corresponding spike in the Br\"unt-V\"ais\"al\"a frequency produces a complicated mode trapping signature in the period spacings. 
   }
{}

   \keywords{subdwarfs -- stars: evolution  -- stars: oscillation -- methods: numerical}

   \maketitle

\section{Introduction}
Hot B-type subdwarfs are identified as core He-burning stars surrounded by a very thin H-envelope \citep{heber1986}. This places them at the blue extension of the horizontal branch in the Hertzsprung-Russell diagram. Hence, they are also referred to as extreme horizontal branch (EHB) stars. Subdwarf B (sdB) stars are ubiquitous in our Galaxy, where they dominate the population of faint blue objects at high galactic latitudes \citep{green1986}. They are also believed to be responsible for the ultraviolet excess or `UV upturn' in giant elliptical galaxies \citep{brown1997,yi1997}.  A thorough review of hot subdwarfs is given by \citet{heber2009}.

The future evolution of sdB stars is straightforward and undisputed. After He in the core is exhausted, a short phase of He-shell burning follows, and the star may be identified as a hotter subdwarf of O-type (sdO). The H-envelope is too thin to sustain H-shell burning, so the star will end as a C-O white dwarf without evolving through the asymptotic giant branch phase. The formation of sdBs on the other hand, remains a much debated topic. Many formation channels have been proposed such as enhanced mass loss on the RGB \citep{d'cruz1996}, mass loss through binary interaction \citep{mengel1976}, and the mergers of two He white dwarfs \citep{webbink1984}. The relative importance of different formation scenarios has been evaluated by binary population synthesis studies \citep{han2002,han2003}. While these studies are valuable, it should not be forgotten that they are dependent on parametric descriptions of binary interaction. Detailed mass determinations of subdwarf B stars could provide important constraints on the binary interaction mechanism. Accurate astrophysical mass determinations are only possible in special circumstances such as in eclipsing binary systems. Asteroseismology provides an alternative method, since it allows a detailed study of the interior of non-radially pulsating stars. The consistency of both approaches has been achieved for \object{PG\,1336$-$018} (see \citealt{vuckovic2007}, \citealt{hu2007}, and \citealt{charpinet2008}).

Subdwarf B stars exhibit a variety of pulsations. The first sdB pulsator, \object{EC\,14026$-$2647}, was observed by \citet{kilkenny1997} to pulsate in multiple short-period modes. This prototype represents the variable class  \object{V361\,Hya} stars which now includes 42 rapid sdB pulsators, with periods in the range \mbox{80-600 s}. The short-period modes have been interpreted as low radial order, low spherical degree $p$-modes \citep{charpinet1997}.  The \object{V361\,Hya} stars are amongst the hottest sdB stars with effective temperatures between 28 000 K and 35 000 K and surface gravities $5.2<\log g<6.1$. At the cooler end of the EHB, 31 sdB pulsators with periods between 30 min and two hours have been discovered by \citet{green2003}. This variable class is named V1093 Her but is also often referred to as \object{PG\,1716$+$426} stars after the class prototype. The long periods suggest that these stars pulsate in high radial order $g$-modes. Especially interesting are the sdB pulsators that exhibit both $p$- and $g$-mode pulsations. Three of these so-called hybrid pulsators have been found at the intersection of the \object{V361\,Hya} and \object{V1093\,Her} stars  \citep{oreiro2005,baran2005,schuh2006,lutz2008}. 

The opacity mechanism operating in the Fe opacity bump around $\log T=5.3$ has been successfully used to explain the excitation of $p$- as well as the $g$-modes \citep{charpinet1996,fontaine2003}. This opacity bump is caused by Fe accumulation owing to the competing diffusion processes of gravitational settling and radiative levitation. While seismic mass determinations have been achieved for a dozen \object{V361\,Hya} stars using static envelope models (see \citealt{fontaine2008} and references therein), this is not yet the case for \object{V1093\,Her} stars. The reason is twofold. First of all, the $g$-modes have lower amplitudes and longer pulsation periods and thus require higher precision observations and longer observing runs to detect the frequencies. Secondly, envelope models may suffice to describe the \object{V361\,Hya} stars' shallow $p$-modes that probe only the outer layers. However, they are not suitable for modelling $g$-modes that penetrate to deeper stellar regions. Stellar models with detailed information about the He-rich core are required to predict reliable $g$-mode periods.

Another problem for the V1093 Her stars is the discrepancy between the observed and theoretically predicted instability strip. The first theoretical models that show unstable $g$-mode pulsations \citep{fontaine2003} have much lower effective temperatures than observed with a discrepancy of $\sim$ 5000 K. \citet{jeffery2006b} managed to place the blue edge of the instability strip within $\sim$1000 K of the observed blue edge. They did this by using OP opacities (\citealt{badnell2005}) rather than OPAL opacities (\citealt{iglesias1996}), and by enhancing the Ni abundance in the envelope in addition to Fe. Thus, their work suggests that  sophisticated models including up-to-date opacities and diffusion of other Fe-group elements are required to address the instability issue properly. 

An exploratory study of how the pulsational properties of sdB stars are related to their internal structure has been carried out by \citet{charpinet2000}. They showed that the He-H transition zone between the He-rich core and H-rich envelope is responsible for trapping  and confining $g$-modes in sdB stars. This is similar to the well-known mode trapping phenomenon in compositionally stratified white dwarfs \citep{winget1981}. Mode trapping results in deviations from the asymptotic constant period spacing, making the $g$-modes sensitive to the mass of the H-envelope. Furthermore, the C-O/He-transition region between the convective and the radiative part of the core will also influence the $g$-mode period distribution. This may be a new interesting way to follow the He-core evolution, which would also have implications in a broader context for horizontal branch stars. 

Here, we shall explore the sensitivity of the $g$-modes to certain assumptions about the He-flash. This energetic event characterizes the start of He-ignition in degenerate cores of low-mass stars  ($<$2 M$_{\odot}$). Most stellar evolution codes encounter numerical difficulties in calculating this phase, and even in successful cases uncertainties remain because of the approximative treatment of convection. Three-dimensional hydrodynamic calculations are necessary to investigate this enigmatic phase of stellar evolution (e.g., \citealt{deupree1996,dearborn2006,mocak2008}), but these are expensive in computing time. When many models are needed, as in forward modelling, it is more practical to construct post-flash models without following the previous evolution in detail. One should then use the outcome of detailed studies to ensure that the stellar models are as realistic as possible. Most simulations indicate that the initial flash starts off-centre, followed by several mini-flashes moving towards the centre. While it is generally found that most of the inner region will be convectively mixed, there are uncertainties in the outer extent of the convective region and the amount of He-burning during the He-flash varying from 3\% to 7\% in mass fraction \citep{piersanti2004,serenelli2005}. The edge of the He-flash-induced convective region will leave a chemical composition gradient, which we believe can produce additional mode trapping features.

We start with constructing zero age extreme horizontal branch (ZAEHB) models in Sect.~\ref{zaehb}. We make an analytical fit to the H-profile in the core-envelope transition region and examine its influence on the $g$-mode periods (Sect.~\ref{profile}). The effects of uncertainties in the He-flash treatment, i.e.~convective mixing and C-production, are explored in Sect.~\ref{heflash}. In Sect.~\ref{ehb}, we perform a stability analysis on a grid of evolutionary sdB structures during the core He-burning phase. We show the impact of the opacities and H/He-diffusion on the instability strip. The conclusions are summarized in Sect.~\ref{conclusions}.

\section{Zero age extreme horizontal branch models}\label{zaehb}
\subsection{Approach}
The stellar models were calculated using the stellar evolution code STARS (\citealt{eggleton1971, han1994, pols1995}). The details of the input physics used in this work are described in \citet{hu2009}.  By default, we constructed the stellar models with OPAL opacities (except in Sect.~\ref{stability}). The pulsational properties were calculated with the adiabatic oscillation code OSC \citep{scuflaire2008} in combination with the non-adiabatic code MAD by \citet{dupret2001}.

Our procedure of constructing ZAEHB models is as follows. First, we take the stellar structure model of a higher mass (2.25 M$_{\odot}$) model that just ignited helium non-degenerately. We then impose an artificial mass loss ($10^{-6}$ $M_*$ yr$^{-1}$) until the remaining mass is of our choice. Finally, we allow the star to evolve, while slowly changing the chemical abundances to the values we wish (see Sect.~\ref{profile}), and after switching off composition changes produced by nuclear reactions and convective mixing. We ensure that the new ZAEHB model is fully converged by checking the increase in the timestep that the evolution code gives for converged models. 

\subsection{The He-H chemical transition profile}\label{profile}
\begin{figure}[!htp]
   \begin{center}
  \includegraphics[angle=-90, width=9cm]{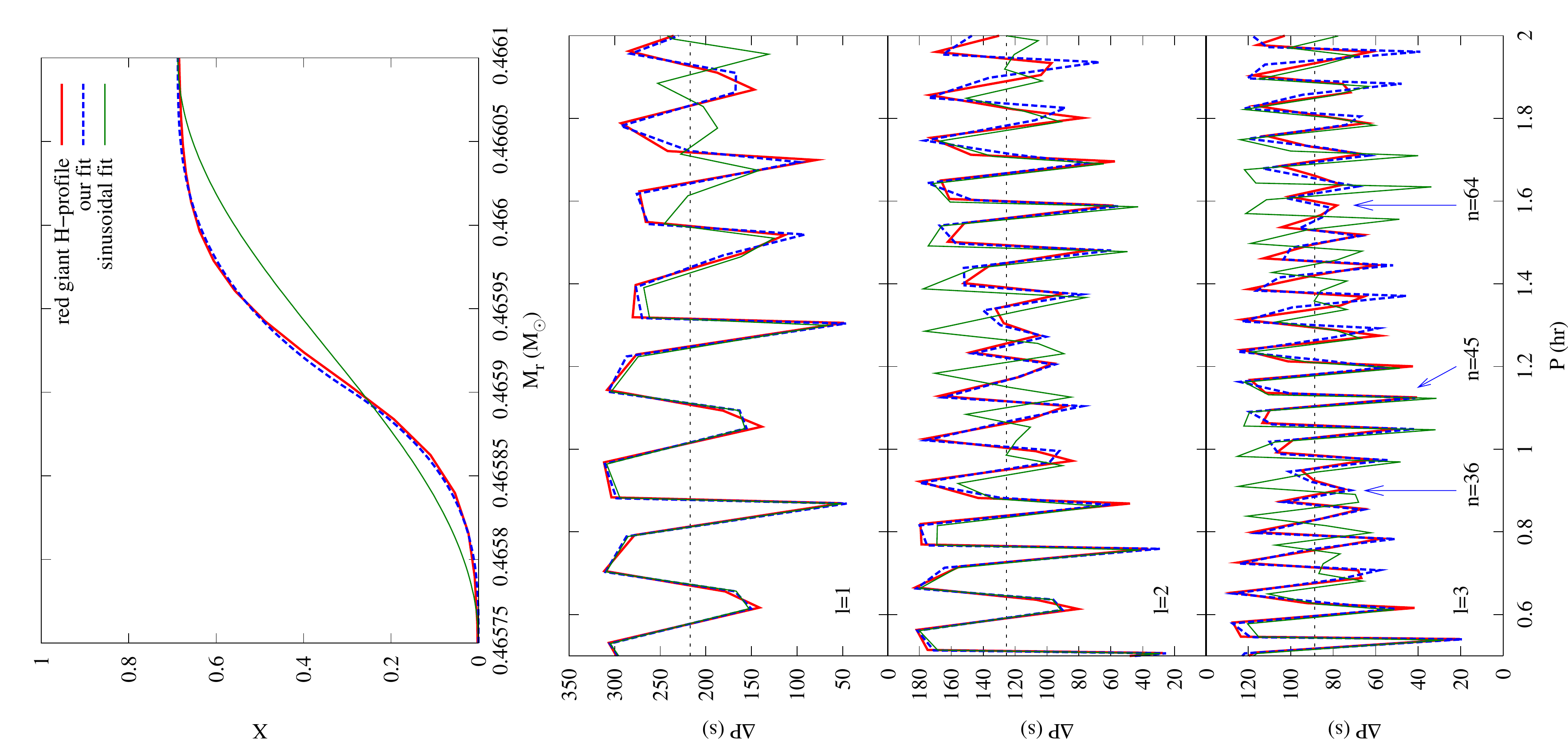}
   \caption{The effect of the H-profile on mode trapping. \emph{Upper panel}: the H-profile in the core-envelope transition zone. We compare the H-profile of a red giant model (thick solid line) to our polynomial Eq.~(\ref{poly}) (blue dashed line) and to the sinusoidal Eq.~(\ref{sinus}) (thin solid line). The three \emph{lower panels} give the period spacing between modes of increasing radial order $n$ as a function of the pulsation period for the modes of spherical degree $l=1,2,3$. We indicated with the horizontal dotted lines the asymptotic constant period spacing.
}
              \label{artprof}
              \end{center}
    \end{figure}

Commonly, one constructs ZA(E)HB models without following the previous evolution. However, special care must be taken when reconstructing the stellar chemical abundances. In particular, the shape of the H-profile in the core-envelope transition zone is very important to $g$-modes because of the associated mode trapping. Therefore, we first evolved a $1.00$ M$_{\odot}$ ZAMS model to the tip of the red giant branch and made a careful fit to  the H-profile of the red giant (see top panel of Fig.~\ref{artprof}). We suggest a formula that fits the H-profile very well using two third degree polynomials:
\begin{eqnarray}\label{poly}
X(M_r) & = & 0 \hspace{25ex}  \textrm{for }M_r<M_{1},\\
&=&(A(M_r-M_{1}))^3 
\hspace{13ex} \textrm{for }M_1\leq M_r < M_{\rm link},\nonumber\\
& = & (B(M_{r}-M_2))^3 + X_{\rm surf}\hspace{6.5ex}
\textrm{for }M_{\rm link}\leq M_r \leq M_2, \nonumber \\
 & = & X_{\rm surf} \hspace{22ex} \textrm{for }M_r > M_2. \nonumber            
\end{eqnarray}
We found that the values $M_1=0.46575$ M$_{\odot}$, $M_2=0.4661$ M$_{\odot}$, $M_{\rm link}=M_1+1.5\times 10^{-4}$ M$_{\odot}$, $X_{\rm surf}=0.688$, and  $A=4400$ provide the best-fit profile. The coefficient $B$ is then determined by assuming that the curve is continuous at $M_r=M_1+M_{\rm link}$.

We compared our approximation to a sinusoidal profile as proposed by \citet{sweigart1976} and used in the ZAHB models of \citet{jeffery2006a}.  The H abundance in the transition zone according their approximation is given by:
\begin{equation}\label{sinus}
X (M_r) = 0.5X_{\rm surf}\Big[1-\cos \Big(\pi\frac{M_r-M_1}{M_2-M_1}\Big) \Big] \quad\textrm{for }M_1\leq M_r \leq M_2. 
\end{equation}
In the top panel of Fig.~\ref{artprof}, we show both approximations, and find that our fit provides a more accurate representation of the red giant's H-profile.

We constructed three ZAEHB models of total mass 0.470 M$_{\odot}$, i.e., with a H-rich envelope of $9\times10^{-4}$ M$_{\odot}$ above the He-H transition zone. In model (i), the abundances were taken from the red giant model throughout the star. In models (ii) and (iii), the H-profile is given by our suggested Eq.~(\ref{poly}) and by the sinusoidal Eq.~(\ref{sinus}), respectively. In the latter two cases, we simply assume that $Z=0.02$ and there is a \citet{grevesse1993} metal mixture throughout the star. In model (i), there are minor effects on the metallicity because of CNO cycling, atomic diffusion, and convective mixing in the previous evolution. We note that we have not yet accounted for flash-induced C-production or mixing. This is to ensure that our comparison is meaningful here. The effects of the He-flash will be evaluated separately  in Sect.~\ref{flash}.

It is well known that in the  asymptotic limit the $g$-mode periods are equally spaced in the radial order $n$ \citep{tassoul1980}. However, sharp features in the Br\"unt-V\"ais\"al\"a frequency cause mode trapping, which manifests itself in deviations from the uniform period spacing. For an in-depth treatment of mode trapping in white dwarfs, sdB stars, and main-sequence stars,  we refer the interested reader to e.g., \citet{brassard1992}, \citet{charpinet2000}, and  \citet{miglio2008}, respectively. Here, it is important to realize that the period spacing  between two consecutive modes in radial order ($\Delta P_l=P_{l, n+1}-P_{l,n}$) 
can be described by an oscillatory component superposed on the asymptotic constant period spacing. The periodicity $\Delta n$ of the oscillatory component is related to the \emph{location} of the chemical gradient, while its amplitude is determined by the \emph{steepness} of the gradient. Trapped modes are then located at the minima in the $\Delta P$ versus $P$ diagram. From the asymptotic theory \citep{tassoul1980}, one can derive the approximative relations
\begin{equation}\label{deltak}
\Delta n=\frac{\Pi_{\rm H}}{\Pi_0}
\qquad\textrm{or}\qquad
P_{l,i+\Delta n}-P_{l,i}=\frac{\Pi_{\rm H}}{\sqrt{l(l+1)}},
\end{equation}
where
\begin{equation}\label{pi}
\Pi_{\rm H}=2\pi^2\Big(\int^R_{r_H} \frac{|N|}{r}dr\Big)^{-1}
\quad\textrm{and}\quad
\Pi_{0}=2\pi^2\Big(\int^R_{r_c} \frac{|N|}{r}dr\Big)^{-1}.
\end{equation}
The Br\"unt-V\"ais\"al\"a (BV) frequency $N$ is given by
\[
N^2=\frac{GM_r}{r^2}\Big(\frac{1}{\Gamma_1}\frac{d \ln P}{d r}-\frac{d \ln \rho}{d r}\Big).
\]
The integration boundary $r_{\rm H}$ indicates the location of the He-H transition zone, $r_c$ the edge of the convective core ($=0$ for a ZAEHB model), and $R$ is the total stellar radius. The asymptotic \emph{constant} period spacing is given by $\Pi_0/\sqrt{l(l+1)}$. 

Figure \ref{artprof} shows the sensitivity of the $g$-modes to the exact shape of the H-profile. We plotted the period spacing $\Delta P$ against the pulsation period $P$, for the modes $l=1,2,3$. We see that our polynomial fit to the H-profile reproduces the $g$-mode spectrum of model (i) very well. In contrast, the sinusoidal H-profile leads to a significantly different mode trapping signature. The deviations become more distinct at higher spherical degree $l$ and higher radial order $n$. Another interesting feature can be seen especially for $l=3$ in the $\Delta P$ versus $P$ diagrams, where there are two oscillatory components in the period spacing. We see a short periodicity in the radial order of $\Delta n_1=3$, modulated by a long periodicity of $\Delta n_2=28$. This is at first instance unexpected, since we only have one chemical transition zone in our ZAEHB models. 

Numerical integration of Eq.~(\ref{pi}) for our model (ii) gives $\Delta n=3$, which accounts for the short periodicity. The long periodicity, however, cannot be explained by the above relations. To understand this effect, one should realize that trapping occurs for a mode if the nodes of its eigenfunctions are suitably located with respect to the BV-discontinuity. In other words, only modes with a certain phase with respect to the discontinuity are trapped, which gives rise to the periodicity of Eq.~(\ref{deltak}).  In reality, the BV-discontinuity of course has a finite width, which was not accounted for in the derivation of Eq.~(\ref{deltak}). Therefore, depending on the change of phase \emph{within} the BV-discontinuity, certain modes can be trapped more efficiently than others, leading to the second periodicity. Again from the asymptotic theory of \citet{tassoul1980}, but now taking into account the finite width of the BV-spike, it can be shown that approximately (Hu et al., in preparation),
\begin{equation}\label{k2}
\Delta n_2=\frac{\Pi_{\rm H2}}{\Pi_0}
\qquad\textrm{with}\qquad \Pi_{\rm H2}=2\pi^2\Big(\int_{r_1}^{r_2}\frac{|N|}{r}dr\Big)^{-1},
\end{equation}
where $r_1$ is the radius at the inner boundary of the BV-spike and $r_2$ is at the outer boundary. Numerical integration of Eq.~(\ref{k2}) for our model (ii) indeed results in $\Delta n_2=28$, showing that the approximate relations hold up very well. 

\begin{figure}[!htp]
   \begin{center}
  \includegraphics[angle=-90, width=9cm]{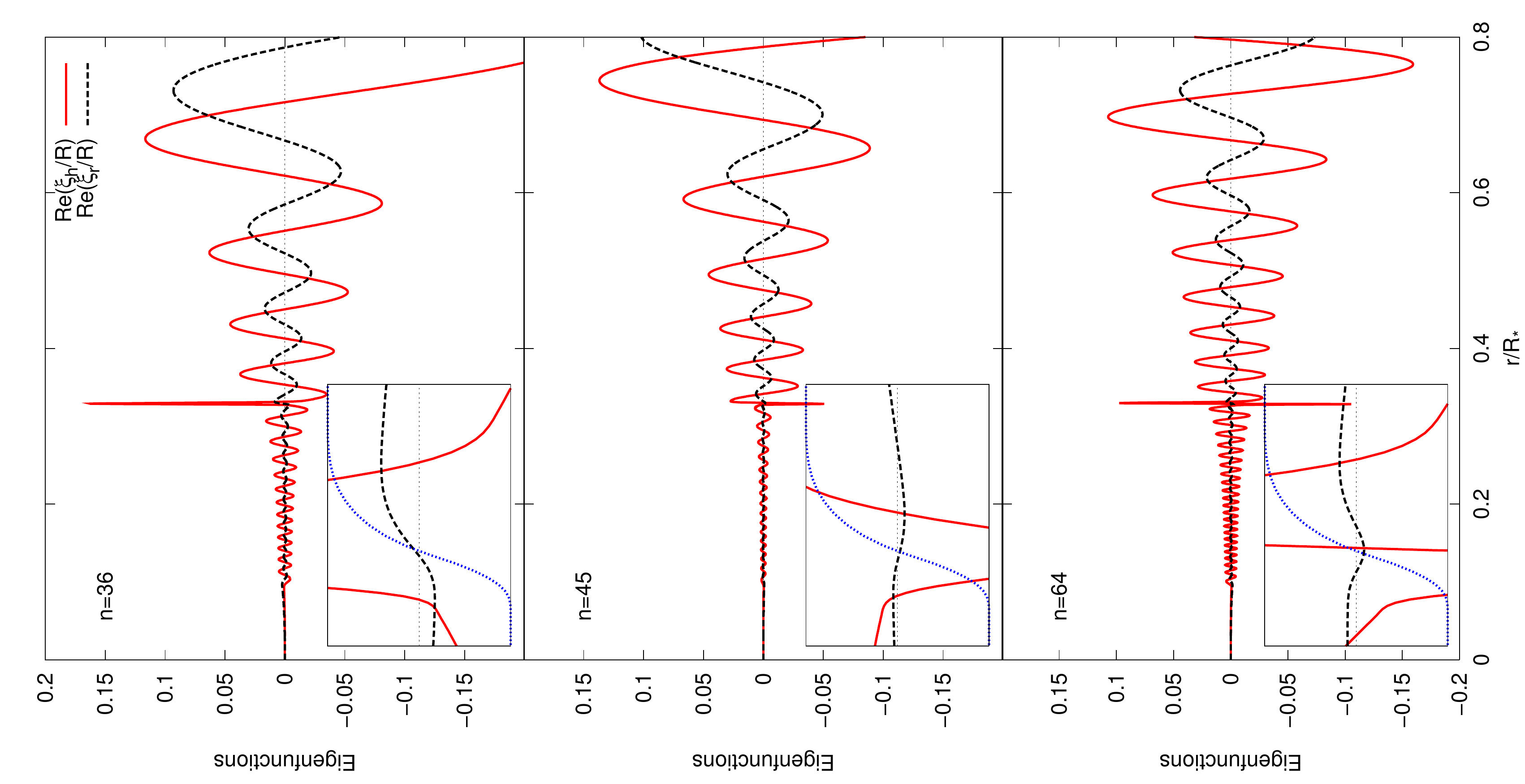}
   \caption{Normalized displacement eigenfunctions $\xi_r$ (dashed line) and $\xi_h$ (solid line) as a function of the normalized stellar radius for ZAEHB model (ii). The \emph{upper panel} is for mode $(l,n)=(3,36)$, the \emph{middle panel} for $(l,n)=(3,45)$, and the \emph{lower panel} corresponds to $(l,n)=(3,64)$ (see Fig.~\ref{artprof} bottom panel). We zoom into the He-H transition zone to show the behaviour of the eigenfunctions in the $g$-mode cavity. In the close up, we also plot the H-profile (dotted line) to indicate the extent of the transition zone. } 
              \label{eigen}
              \end{center}
    \end{figure}

To clarify this, it is helpful to examine the displacement eigenfunctions,  $\xi_r$ (radial) and $\xi_h$ (horizontal). We find that a mode is trapped when it has a node in $\xi_h$ at the bottom of the He-H transition zone directly followed by a node in $\xi_r$. This is consistent with the findings of \citet{brassard1992} for white dwarfs. 
In Fig.~\ref{eigen}, we plot for model (ii) the displacements for three trapped modes of radial order $n=36$, $n=45$, and $n= 36+\Delta n_2=64$. These have a small, large, and again small amplitude in $\Delta P$, respectively (see Fig.~\ref{artprof}). We see that mode $n=36$ also has a node in $\xi_h$ at the top of the transition zone, while mode $n=45$ has a node there in $\xi_r$.
We find this to be generally true and conclude that modes are more efficiently trapped when the first node encountered in the He-H transition zone corresponds to $\xi_h$ and the last node to $\xi_r$. After $\Delta n_2$ modes, the nodes are again suitably placed. This can be seen by comparing $n=36$ and $n=64$: the location of the nodes of $\xi_h$ with respect to the boundaries is approximately the same, while for $n=64$, there is  one additional node in-between. The interpretation of this interesting phenomenon will the subject of a forthcoming paper (Hu et al., in preparation).

\subsection{The helium core flash}\label{heflash}
   \begin{figure*}[!htp]
   \begin{center}
  \includegraphics[angle=-90, width=18cm]{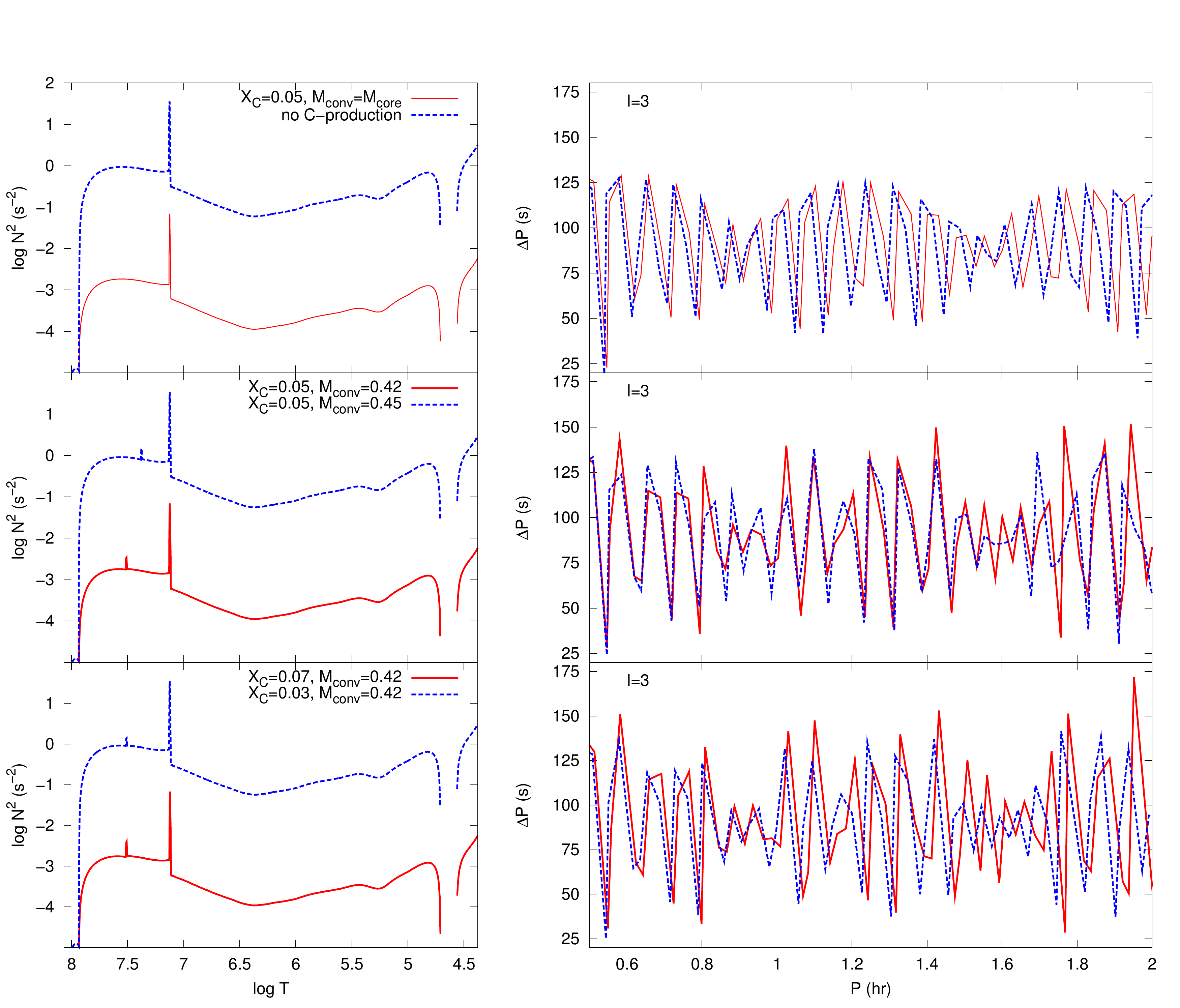}
   \caption{Effects of different assumptions about the He-flash on the $g$-mode spectrum of ZAEHB models. \emph{Left}: the BV-frequency throughout the star in terms of the temperature. The dashed curves are shifted upwards with $\log500$ for visibility. \emph{Right}: the period spacing between consecutive g-modes as a function of the pulsation period. In the \emph{upper panels} the case of C-production during the He-flash of 5\% throughout the entire He-core (solid line) is compared to the case of no C-production (dashed line). The \emph{middle panels} show the results for ZAEHB models where convective mixing was assumed up to $M_r=0.42$ M$_{\odot}$ (solid line) and $M_r=0.45$ M$_{\odot}$ (dashed line). In the \emph{lower panels} the amount of C-production is varied from 3\% (dashed line) to 7\% (solid line).} 
              \label{flash}
              \end{center}
    \end{figure*}
   \begin{figure*}[!htp]
   \begin{center}
  \includegraphics[angle=-90, width=18cm]{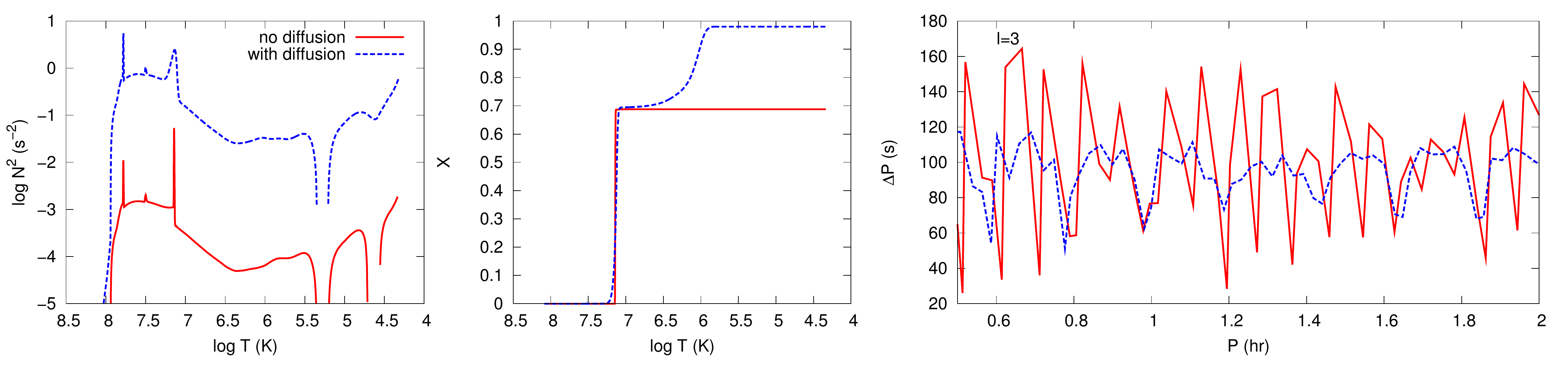}
   \caption{Comparison of sdB models with (dashed line) and without (solid line) atomic diffusion at an EHB age of $5\times10^7$ years. The \emph{left panel} shows the  BV-frequency as a function of the stellar interior temperature. The dashed curve is shifted upwards with $\log500$ to avoid overlapping. The three BV-spikes from left to right correspond to 1) the boundary of the convective core, 2) the edge of the He-flash-induced mixing region (small spike), and 3) the He-H transition zone. The \emph{middle panel} shows the effect of diffusion on the H-profile.  The \emph{right panel} shows the period spacing versus the period, illustrating the mode trapping caused by the chemical transitions. } 
              \label{diff}
              \end{center}
    \end{figure*}

In previous models we did not account for the inner core abundances being in reality affected by nuclear burning and convective mixing during the He-flash. From now on, we treat the He-flash in a `standard' way by converting 5\% of the He-mass fraction into C and assume that the He-flash-induced convection region extends to 90\% of the He-rich core (i.e., $M_r=0.42$ M$_{\odot}$) as suggested by 3D hydrodynamical simulations of \citet{dearborn2006}. To test the influence of nuclear burning and convective mixing during the He-flash, we also constructed post-flash ZAEHB models where i) C is increased by 3\% and 7\%, respectively, and ii) the core is convectively mixed with up to 97\% ($0.45$ M$_{\odot}$) and 100\% (0.46575 M$_{\odot}$) of the core, respectively.

Figure \ref{flash} shows the effects of the different assumptions about the He-flash on the $g$-mode spectrum. The upper panels indicate that increasing C in the entire He-core does not have a significant effect. There is only a minor shift to higher pulsation periods because of a slight increase in the stellar radius, but the $g$-mode spectrum remains qualitatively the same. However, if the He-flash-induced C-production is not mixed throughout the entire core, there is an additional chemical gradient, and thus an additional spike in the  BV-frequency. This results in a complicated mode trapping pattern in the period spacings, that depends on the location of this spike, as shown in the middle panels of Fig.~\ref{flash}. In the lower panels, one can see that the strength of the BV-spike, and thus the effect of mode trapping, is sensitive to the amount of C produced during the He-flash.

We emphasize that the ZAEHB models do not yet have a C-O convective core, and thus the He-H transition and the edge of the He-flash-induced  convective region are mainly responsible for mode trapping in these models. The thin He-convection zone near the surface, which is visible in Fig.~\ref{flash} as negative $N^2$ around $\log T=4.7$, has only a very small effect on  the $g$-modes.

\section{Evolutionary extreme horizontal branch models}\label{ehb}
\subsection{Approach}
During the EHB evolution, Fe accumulates in the driving region (around $\log T=5.3$) due to radiative levitation \citep{charpinet1997, fontaine2006}.  It is reasonable to expect that other Fe-group elements, such as Ni, are affected similarly by radiative levitation. Considering the work of \citet{jeffery2006b}, we included both Fe and  Ni enhancement in our sdB models. We built tables from OPAL opacities \citep{iglesias1996}, where the number fractions of Fe and Ni are enhanced by a factor $f=1, 2, 5, 10$ relative to the \citet{grevesse1993} metal mixture. Of course, the number fractions of the other metals are accordingly reduced. The OPAL radiative opacities are supplemented by low-temperature opacities \citep{ferguson2005} and conductive opacities \citep{cassisi2007}. 

We increased the enhancement factor $f$ during the EHB evolution from 1 to 10 with a Gaussian centred around $\log T=5.3$. Already after $10^7$ yr, we have $f(\log T=5.3)\approx8$ (see Fig.~1 of \citealt{hu2008}). The opacity is then calculated by interpolating between the opacity tables for different $f$. We note that Fe/Ni-accumulation is only accounted for in the opacities. This is reasonable since the mass fractions of Fe and Ni are low even after enhancement, and these elements are not involved in nuclear reactions. Thus, the only relevant effect of the enhancement is in the stellar opacities.
We follow the approach by \citet{hu2008}, but with two noteworthy differences taking into account the work of \citet{jeffery2006b}. First of all, as we have already mentioned, Ni is enhanced in addition to Fe. Secondly, the mass fractions of the non-enhanced metals remain constant resulting in a slight increase in the total metallicity, whereas in our previous work, the other metals were reduced to keep the total metallicity constant. 

\subsection{Atomic diffusion}\label{diffusion}
   
Our models include H/He-diffusion as described in \citet{hu2009}. 
In the presence of diffusion, steep composition gradients are broadened leading to less efficient mode trapping. This can be seen in Fig.~\ref{diff}, where we show the same ZAEHB model evolved with and without atomic diffusion. Both models have an EHB age of $5\times10^7$ years. As expected, the amplitude of the oscillatory component in the period spacing is much smaller in the case of diffusion. Because of the lower amplitude, the effect of $\Delta n_2$ ($\approx 30$ in this case) is less significant, although still visible. The mode trapping pattern is at this stage even more complicated, because the sdB star has developed a convective core, which leads to another BV-discontinuity (see left panel of Fig.~\ref{diff}).

In Sect.~\ref{profile}, we detected the extreme sensitivity of the $g$-mode spectrum to the shape of the H-profile. This shows that special care must be taken when modelling the He-H transition. However, as atomic diffusion tends to wash away the differences in the initial chemical profiles, the significance of modelling the H-profile very precisely becomes less for evolved models. 

Another interesting effect of He-settling is the disappearance of the He-convection zone, as can be seen in the left panel of Fig.~\ref{diff}. The negative $N^2$ around $\log T =5.3$ corresponds to the Fe surface convection zone. We note that the Fe convection zone was not present in our ZAEHB models, since our models accumulate Fe \emph{during} the evolution. It is not expected that the surface convection zones have a significant effect on the $g$-mode periods because of the small contributions of the surface layers to the weight function (see Fig.~9 of \citealt{charpinet2000}).

\subsection{Stability analysis of a grid of sdB models}\label{stability}
The previous models all have a mass of $0.470$ M$_{\odot}$. For a stability analysis, we constructed models of varying mass so that the observed \object{V1093\,Her} $\log g-\log T_{\rm eff}$ region is mapped. We computed sdB models with a fixed core mass defined as before at 0.46575 M$_{\odot}$, and envelope masses: $2.5\times 10^{-4}$, $5.5\times 10^{-4}$, $1.25\times 10^{-3}$,  $2.25\times 10^{-3}$,  $4.25\times 10^{-3}$,  $6.25\times 10^{-3}$, and $8.25\times 10^{-3}$ M$_{\odot}$. To make the models as realistic as possible, we took the abundances from the red giant model at the tip of the RGB. The inner 0.42 M$_{\odot}$ of the core is assumed to be mixed during the He-flash and C production is taken to be 5\%.

 It is well-known that mode excitation is very sensitive to the stellar opacities. Thus, we also constructed opacity tables with OP opacities using OP server \citep{badnell2005,seaton2005}. For a fair comparison, we obtained OP tables for the same metal mixture as OPAL (i.e., \citealt{grevesse1993}). 
Because the maximum temperature for OP opacities is $\log T=8$, we use OPAL opacities for He-burning regions that are enriched in C and O.

\begin{figure}[!htp]
   \begin{center}
  \includegraphics[angle=-90, width=9cm]{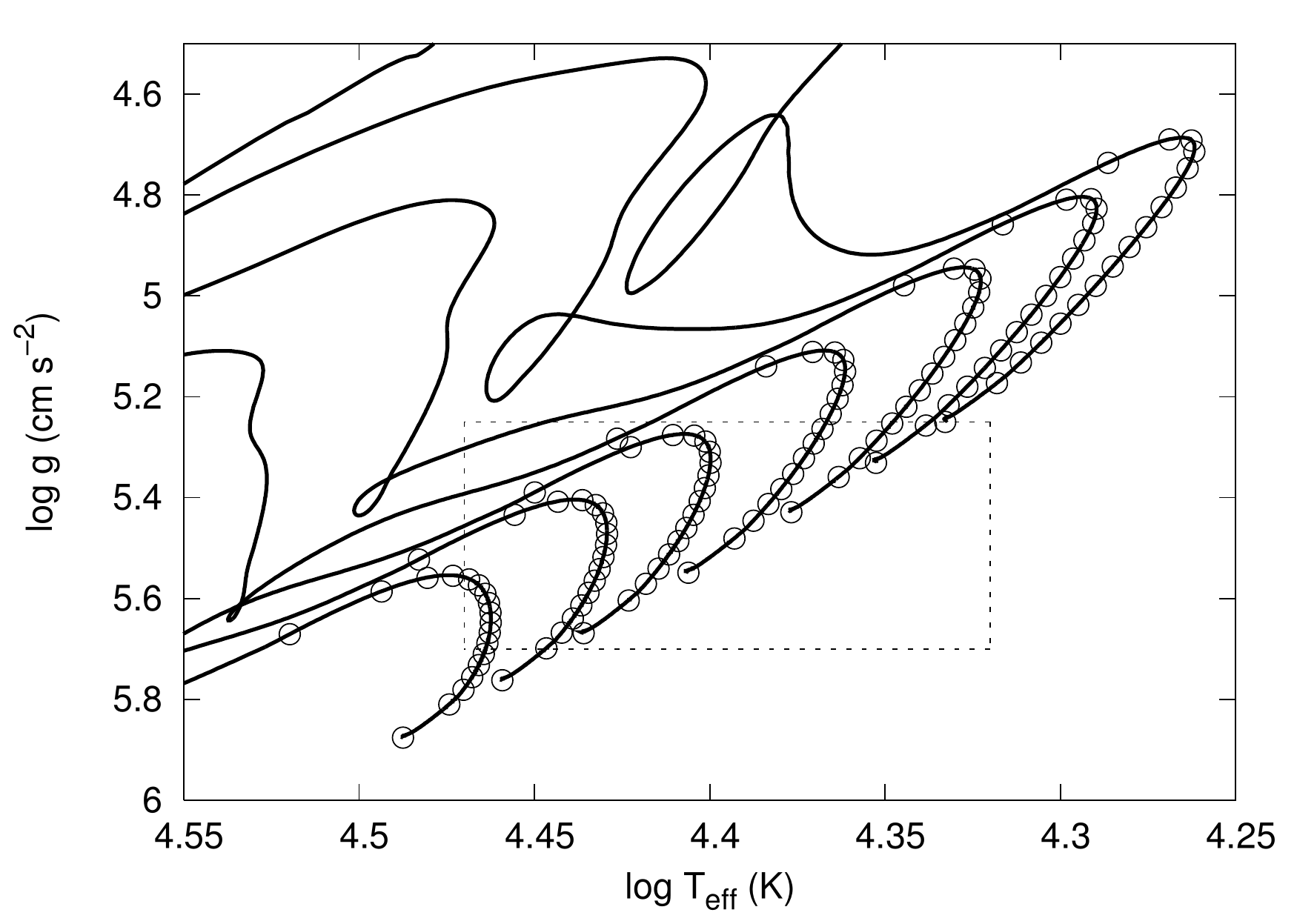}
   \caption{ Evolutionary tracks of our sdB models in the $\log g-\log T_{\rm eff}$ diagram.  The tracks from high to low $T_{\rm eff}$ correspond to sdB masses increasing from 0.466 to 0.474 M$_{\odot}$. It is assumed that 5\% of C is produced during the He-flash and the inner region is convectively mixed up to $M_r=0.42$ M$_{\odot}$. The circles indicate the models used in the stability analysis. The dotted box shows the region in which V1093 Her stars have been observed.} 
              \label{evtracks}
              \end{center}
    \end{figure}

Figure \ref{evtracks} shows the evolutionary tracks computed with OP opacities in the $\log g-\log T_{\rm eff}$ diagram. The tracks for OPAL overlap with these, thus the differences between OP and OPAL opacities have a negligible effect on the global evolution.
The structure models for which we calculated non-adiabatic pulsational properties, are represented by circles. The time interval between the pulsation models is $10^7$ years. 

\begin{figure*}[!htp]
   \begin{center}
  \includegraphics[angle=-90, width=18cm]{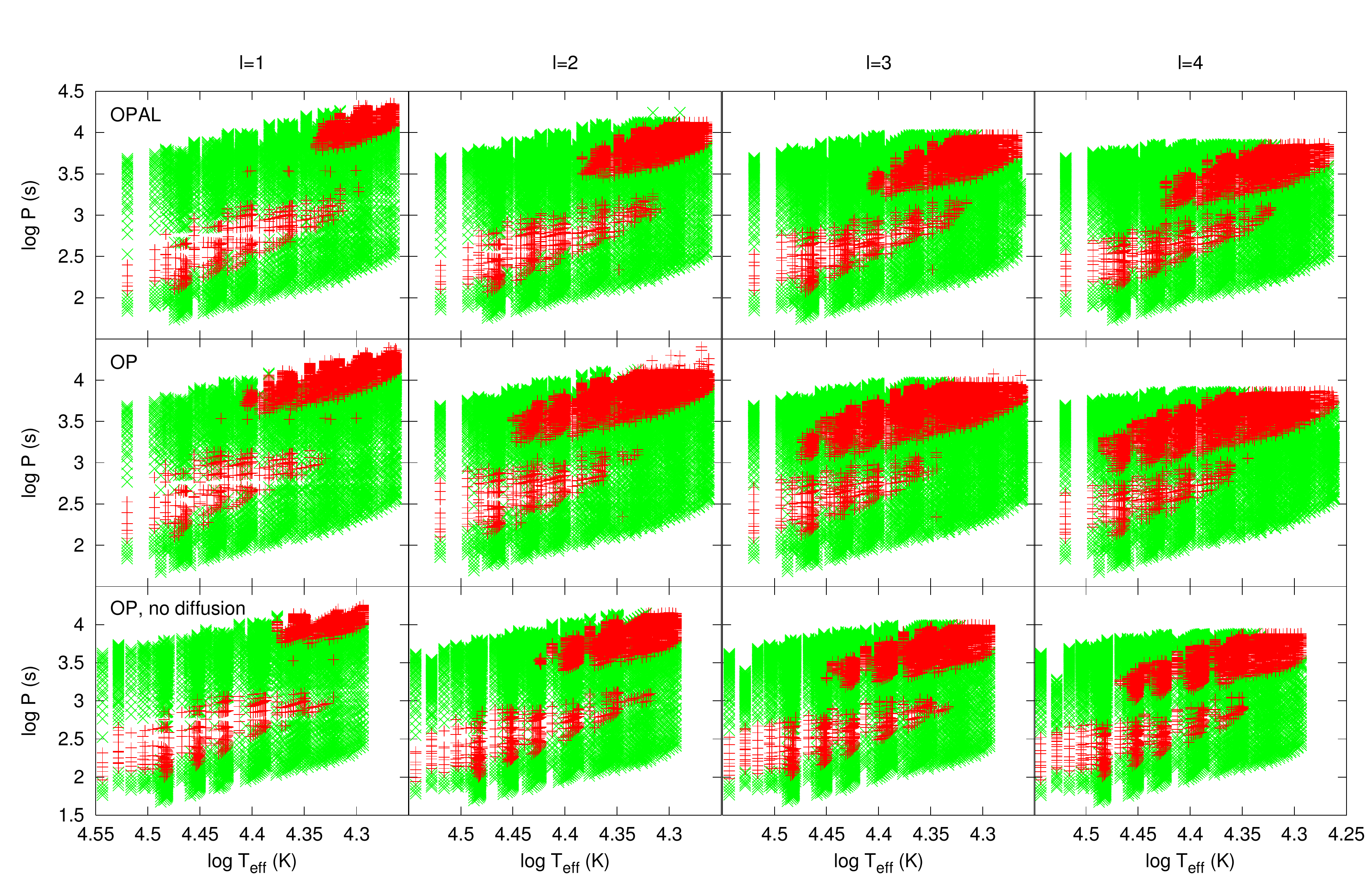}
   \caption{Pulsation periods of stable (green $\times$) and unstable (red $+$) modes as a function of the effective temperature. The \emph{upper panels} are for sdB models computed with OPAL opacities and the \emph{middle panels} are for OP opacities. In the \emph{lower panels} we show the results for OP opacities if H/He-diffusion is not included. In all three cases, the abundances of Fe and Ni are increased around $\log T=5.3$ up to a factor 10 in number during the evolution.} 
              \label{istrip}
              \end{center}
    \end{figure*}

Figure \ref{istrip} shows the instability strip of our models obtained using OPAL opacities (upper panels) and OP opacities (middle panels). In agreement with previous studies (\citealt{jeffery2006b}, \citealt{miglio2007}), we find that with OP opacities, unstable modes are found in hotter stars than with OPAL opacities. This is related to differences in the strength and location of the Fe-group opacity bump as a function of the temperature. For OP opacities, this bump is slightly larger and occurs at higher temperatures (see \citealt{seaton2004}). The thermal relaxation time ($\tau_{\rm th}=c_V T\Delta m/L_*$) in the driving region, i.e., around the opacity bump, is then large enough for efficient driving of the $g$-modes. This is because, first of all, the opacity driving mechanism works most efficiently on modes with periods comparable to $\tau_{\rm th}$. Secondly, only high-order $g$-modes (with long periods) can be excited, for low-order $g$-modes are affected by significant radiative damping. The same phenomenon occurs in SPB stars (\citealt{dziembowski1993, dupret2008}).  

Interestingly, the blue edge of the $l\leq3$ instability strip of our sdB models with OP opacities corresponds to the observed blue edge of the V1093 Her variables at $\log T_{\rm eff}=4.47$.  
Thus, we have identified that atomic diffusion represents another step towards solving the `blue edge problem', in addition to the use of OP opacities and Fe/Ni enhancement suggested by \citet{jeffery2006a, jeffery2006b}.  The closer match between the theoretical and observational instability strips is caused by the settling of He and the outward diffusion of H. As a result, the envelope's mass density decreases and the opacity bump at $\log T=5.3$ moves inwards within the star because of the relation $|\rm{d} T/\rm{d}r| \propto \kappa \rho$. This is towards regions that have a longer thermal relaxation time, and as explained before, modes with longer pulsation periods can now be efficiently driven. These are high-order $g$-modes for which no significant radiative damping occurs. 
For a grid that was evolved without H/He-diffusion and with OP opacities (lower panels of Fig.~\ref{istrip}), we indeed find the same blue edge as \citet{jeffery2006b}, i.e., $\log T_{\rm eff}=4.45$ for $l=3$ modes. However, Ni enhancement remains an essential ingredient, as we checked for models with only Fe enhancement (see Table \ref{blue edge}). Observations confirm the effect of settling, because most sdB stars are He-deficient with $\log [N(\textrm{He})/N(\textrm{H})]\sim-2$ \citep{saffer1994, heber2004}. However, the effect of He-settling is probably less efficient in reality than in our models because of counteracting processes such as weak stellar winds \citep{unglaub2001}, turbulent mixing, and radiative levitation (although the effect of radiative levitation on He is found to be small, see \citealt{michaud2008}).

\begin{table}
\begin{minipage}{\columnwidth}
\caption{Blue edge in terms of $\log T_{\rm eff}$ of the $g$-mode instability strip for our sdB models. \protect\footnote{
The results for the first three assumptions are also shown in Fig.~\ref{istrip}. 
} }
\label{blue edge}
\begin{center}
\begin{tabular}{llllllll}
\hline
\vspace{-0.25cm}\\
Fe&  Ni & H/He-&opacity &\multicolumn{4}{c}{$l$}\\
 & &diffusion & & 1 & 2 & 3 & 4  \\
\vspace{-0.25cm}\\
\hline
\vspace{-0.25cm}\\
yes & yes &  yes &OPAL  & 4.34  & 4.38 & 4.41 & 4.43    \\
yes & yes &  yes & OP      & 4.41 & 4.45 & 4.47 & 4.48  \\
yes & yes &  no &  OP      & 4.38  & 4.43 & 4.45 & 4.47   \\
yes & no &  yes &   OP    & 4.35  & 4.39 & 4.42 & 4.44  \\
\hline
\end{tabular}
\end{center}
\end{minipage}
\end{table}


\section{Conclusions}\label{conclusions}
The precise details of chemical transitions in subdwarf B stars have a considerable influence on the period spacings of high order $g$-modes. This is a signature of mode trapping that occurs due to discontinuities in the BV-frequency. We have shown that the structure of mode trapping depends sensitively on the exact shape of the H-profile in the core-envelope transition zone. Thus, when constructing ZAEHB models without following the previous evolution, the chemical profiles must be modelled carefully. Since the action of mode trapping is so sensitive to the shape of the chemical transition, it is necessary to include atomic diffusion during the evolution, i.e., the processes of gravitational settling, temperature and concentration diffusion. 

Furthermore, if the sdB star started He-fusion in a run-away flash, the inner part of the core is convectively mixed and C is produced up to $\sim$7\%.  The edge of the He-flash-induced convective core is accompanied by a chemical transition.  We showed that this can cause additional mode trapping features resulting in a complicated behaviour of the period spacings. 
Multiple composition gradients lead to simultaneous mode trapping in different regions. The $g$-mode spectra become very complicated, and it will be difficult to directly infer detailed information about the composition gradients from observed $g$-modes. It is therefore crucial to evaluate the importance of different physical processes, and to account for all of them realistically, as we do here.

Finally, and perhaps most importantly, we report on a possible step forward in solving the `blue edge problem' of the V1093 Her instability strip. Our evolutionary sdB models with (i) OP opacities, (ii) the inclusion of H/He  diffusion, and (iii) a parametric Fe/Ni enhancement, show unstable $l\leq 3$ $g$-modes for effective temperatures up to $30,000$ K. This is only achieved when all three ingredients are included in the computations. It should be noted that it remains unclear which $l$-values the observed long-period pulsation modes have. 
Thus, although we have demonstrated the importance of H/He diffusion, solving the blue edge problem effectively will require mode identification, in addition to a realistic treatment of radiative levitation and other transport mechanisms. 
 
\begin{acknowledgements}
HH acknowledges a PhD scholarship through the ``Convenant
Katholieke Universiteit Leuven, Belgium -- Radboud Universiteit Nijmegen, the
Netherlands''. GN is supported by NWO VIDI grant 639.042.813. 
The research leading to these results has received funding from the
European Research Council under the European Community's Seventh Framework Programme (FP7/2007--2013)/ERC grant agreement n$^\circ$227224 (PROSPERITY), as well as from the Research Council of K.U.Leuven grant agreement GOA/2008/04. 

\end{acknowledgements}

\bibliographystyle{aa}
\bibliography{12699}
\end{document}